\documentclass[letterpaper, 10 pt,conference]{IEEEtran}
\usepackage[letterpaper, left=0.75in, right=0.75in, bottom=0.75in, top=0.75in]{geometry}
\usepackage[pdftex]{graphicx}
\usepackage{subfigure}
\usepackage{xcolor}
\IEEEoverridecommandlockouts 
\usepackage[utf8]{inputenc}
\usepackage{graphicx}
\usepackage{subfigure}
\usepackage{cite}
\usepackage{amsmath,amssymb,amsfonts}
\usepackage{algorithmic}
\usepackage{makecell}
\usepackage{url}
\usepackage[justification=centering]{caption}
\usepackage{booktabs}
\usepackage{multirow}
\usepackage{textcomp}
\usepackage{comment}
\usepackage{balance}
\def\BibTeX{{\rm B\kern-.05em{\sc i\kern-.025em b}\kern-.08em
T\kern-.1667em\lower.7ex\hbox{E}\kern-.125emX}}
\begin{document}
\title{\vspace{0.25in}Estimating the Number of HIV+ Latino MSM Using RDS, SS-PSE, and the Census}

\author{Nicholas Budzban,
        Katherine Silverio, 
        and~John Matta
        
        \thanks{Nicholas Budzban, Katherine Silverio and John Matta are with the Computer Science Department, Southern Illinois University Edwardsville, Edwardsville, IL 62026 USA. Corresponding email:jmatta@siue.edu}
        }

\markboth{Journal of \LaTeX\ Class Files,~Vol.~14, No.~8, August~2015}%
{Shell \MakeLowercase{\textit{et al.}}: Bare Demo of IEEEtran.cls for IEEE Journals}

\maketitle

\begin{abstract}
This paper presents a method for estimating the overall size of a hidden population using results from a respondent driven sampling (RDS) survey. We use data from the Latino MSM Community Involvement survey (LMSM-CI), an RDS dataset that contains information collected regarding the Latino MSM communities in Chicago and San Francisco. A novel model is developed in which data collected in the LMSM-CI survey serves as a bridge for use of data from other sources. In particular, American Community Survey Same-Sex Householder data along with UCLA's Williams Institute data on LGBT population by county are combined with current living situation data taken from the LMSM-CI dataset. Results obtained from these sources are used as the prior distribution for Successive-Sampling Population Size Estimation (SS-PSE) - a method used to create a probability distribution over population sizes. The strength of our model is that it does not rely on estimates of community size taken during an RDS survey, which are prone to inaccuracies and not useful in other contexts. It allows unambiguous, useful data (such as living situation), to be used to estimate population sizes. 
\end{abstract}


\IEEEpeerreviewmaketitle

\section{Introduction}
 \IEEEPARstart{D}{etermining} the size of a hard-to-reach or hidden population is of immense importance when planning for health interventions, estimating their success, or budgeting for costs associated with a disease. This is particularly true for a disease like HIV which affects minority populations disproportionately \cite{hess2018diagnoses} and is likely to remain unreported due to stigma surrounding both men who have sex with men (MSM) populations and HIV. Respondent driven sampling (RDS) \cite{heckathorn1997respondent} is a survey technique in which members of a hidden population recruit other members into the survey. This technique has been used not only with HIV populations, but also to collect data on hidden populations such as MDMA users \cite{wang2005respondent} and migrant workers \cite{tyldum2014applying}.

Multiplier methods, which compare independent sources of data, are widely used to estimate the size of hidden populations. However, ``multipliers based on different data sources can yield vastly different results" \cite{abdul2014estimating}, and multiplier methods require random sampling, which is difficult or impossible when hidden members prefer to remain hidden.

Another approach to estimating the size of hard-to-count populations is \textit{network scale-up} (NSUM)  \cite{Bernardii11,guo2013estimating,feehan2016generalizing}. NSUM is a simple but powerful idea, where respondents from the hidden population are asked to estimate the number of members $m_i$ of that population in their social circle, as well as to estimate the overall size of their social circle $\hat{c}_i$. This ratio is then scaled up by the size of the overall population $N$ to produce an estimate $\hat{e}$ of the size of the hidden population, according to the formula
\begin{equation*}
    \hat{e} = \frac{\sum_i m_i}{\sum_i \hat{c}_i } \times N.
\end{equation*}

The use of NSUM requires planning when the survey data are collected, such that respondents are asked about the size of their network as well as the size of personally known sub-networks with sought-after characteristics. This requires asking respondents for estimates that may be error-prone. Additionally, the data collected are not useful in any other meaningful sense. The dataset used in this paper does not contain the respondent's overall network size information, so another method is developed. 

We use the Latino MSM Community Involvement (LMSM-CI) dataset \cite{ramirez2013latino} to estimate the population size of Latino MSM with HIV in Cook County, (located in Chicago, Illinois USA), and San Francisco County (located in San Francisco, California, USA). This dataset is the work of Dr. Jesus Ramirez-Valles and was obtained for a study to determine whether community involvement reduced the risk of HIV in minority populations \cite{ramirez2008hiv}. The theoretical framework for that study came from Ramirez-Valles’ 2002 paper \cite{ramirez2002protective}, in which he proposed guidelines for reducing HIV risk behavior.

In 2011 Gile introduced a successive sampling (SS) based estimator for population means that does not require knowledge of the subpopulation and uses data collected through respondent-driven sampling alone \cite{gile2011improved}. This method was improved upon by Handcock, Gile, and Mar \cite{handcock2014estimating}, who implemented their methods in the RDS Analyst software package \cite{handcock2014rds}. The estimation techniques described in this paper use RDS Analyst and its included successive sampling - population size estimation (SS-PSE) package. They are similar to the techniques used by Johnston, McLaughlin, Rouhani, and Bartels in \cite{johnston2017measuring}. In that paper RDS Analyst and the SS-PSE package are used, with experts providing a population estimate for RDS Analyst's posterior distribution tool. Here, however, instead of asking experts for a population estimate, we use a novel combination of data sources, including American Community Survey (ACS) census data and information on LGBT populations from the Williams Institute \cite{WilliamsInstitute}. The single question of living situation, and in particular the response of "Same-Sex Householder" is used as a bridge between the LMSM-CI survey and the corresponding census and Williams Institute data. We show that incorporating this additional data solves the multiplier method problem of requiring random data sources, as well as the NSUM problem of requiring survey-specific network estimation questions. 

\section{Related Work}

Many attempts have been made to estimate the size of hidden populations such as MSM and injecting drug users. One early method by Archibald et al. uses data from HIV testing results and combines it with HIV testing behavior data \cite{archibald2001estimating}. The size estimate for these populations is determined by dividing the number of the studied population (either MSM or injecting drug users) by the proportion of the respective group that reported being tested.

In Livak et al., the authors estimate the population of young Black MSM (YBMSM) living on the south side of Chicago, the area of the city where HIV is most prevalent \cite{livak2013estimating}. They use three methods: an indirect approach, data from the National Survey of Family Growth, and a modified Delphi approach. They determine the crude average of these methods and estimate the population of YBMSM to be 5,578. Wesson, Handcock, McFarland, and Raymond also study African American MSM, but focus on San Francisco. They use the respondents' personal network sizes, collected as part of the survey, with RDS-Analyst to make their estimation \cite{wesson2015if}. The current study is similar to these in that we examine both Chicago and San Francisco. However, instead of concentrating on African American MSM, our target population is Latino MSM with HIV. 


Safarnejad, Nga, and Son estimate the hard-to-reach population of MSM in Ho Chi Minh City and Nghe An province, Vietnam \cite{safarnejad2017population}. The authors use a multiplier method, social application technology, and internet surveys. 
Raymond, McFarland, and Wesson \cite{raymond2019estimated} update the estimated MSM population size in San Francisco \cite{raymond2013estimating} using multiple methods and data sources. The authors obtain this updated hidden population size estimate by using several estimates synthesized by the Anchored Multiplier method (a Bayseian method). The Anchored Multiplier method was developed by Wesson et al. \cite{wesson2018bayesian}. 

In \cite{zhang2007advantages}, the authors combine census and RDS data to estimate the size of female sex workers in a city in western China. The authors determine that census data tends to underestimate population sizes and could be used as a lower limit. They also find that multiplier methods could be used to determine population size estimates for larger geographic regions. 

In \cite{handcock2015estimating}, Handcock, Gile, and Mar study two different hard-to-reach populations in El Salvador. They use methods to estimate these population sizes from recruitment patterns obtained from RDS data. 

In addition to specific population estimation techniques, there is a body of related literature that examines the techniques for strengths and weaknesses, as well as for accuracy. 
Fearon et al. study using multiplier methods, with an emphasis on examining the often large variance in the resulting estimates \cite{fearon2017sample}. To obtain a more confident estimate of a hard-to-reach population when using RDS data, the authors suggest changes to survey collection methods, such as a longer period of coupon distribution, as well as beginning collection with enough seeds to adequately capture the diversity of the hidden population. An RDS survey is analyzed as a graph in \cite{grubb2020identifying}.

Abdul-Quader, Baughman, and Hladik observe in \cite{abdul2014estimating} that estimation methods for crucial populations are inadequate. In their paper, they summarize and review six methods for estimating the size of a key population, including the single-survey method based on an RDS survey.



\section{Methods}

\subsection{Heckathorn's Respondent Driven Sampling}

RDS is a data collection technique where samples are generated from a random walk along nodes in the underlying network with sampling probability proportional to the node's degree \cite{handcock2014estimating}. Heckathorn showed that even though the ``seeds" of the network are chosen by the researchers and may be considered a convenience sample, the subsequent samples chosen by the participants become increasingly independent and disconnected from the seeds and bias of the researchers \cite{heckathorn1997respondent}.
\par
There are several components required for a successful RDS sampling process and its subsequent analysis. First, members of the target population must be able to identify and recruit other members of that population. This internal recruitment is one of the strengths of the RDS process, as members of hidden populations are often stigmatized and wish to keep their identities secret outside of the internal network. RDS allows an individual's participation in the study to be anonymous and unknown to anyone outside of their recruiter and their recruitees, and so RDS seems to be an ethical way to study populations whose individuals wish to remain hidden from those outside.
\par 
As for analysis, there are several assumptions being made for us to consider the collected data as an unbiased probability sample. First, the recruit's recruitment process must be done independently and uncorrelated with the study's variables -- else the analysis of those variables will suffer from sampling bias. Furthermore, RDS assumes that a sample's recruitment probability is the inverse of their network ``visibility" -- the number of people who know them well enough to recruit them. This assumption about sampling probability yields the RDS-II probability estimator, the basis of Gile's SS estimator which we use in this study.


\subsection{Gile's Successive Sampling Population Size Estimation}

Successive Sampling Population Size Estimation (SS-PSE) is a technique developed by Gile and Handcock \cite{handcock2014estimating} to make inference on a network's size given the order and degree of nodes in the random-walk sampled network. The basic idea is that we expect to sample nodes with a higher degree earlier in the random walk process, and the prevalence of large nodes late in the random walk suggests we are only scratching the surface of a large, untapped network. 
The reason this technique is so appealing for both our use and RDS studies in general is because RDS studies collect the SS-PSE required data by default \cite{handcock2014estimating,handcock2015estimating}.

SS-PSE as implemented by RDS Analyst \cite{handcock2014rds} has a few additional requirements. First, users must specify a prior estimate on the network's size in the form of a mean, median, mode, or 50\% confidence interval. This information will be used by RDS Analyst to fit a prior beta distribution to the user's input. Furthermore, RDS Analyst will automatically fit an exponential degree distribution to the sampled network without any additional user input. From there, SS-PSE will undergo a Markov Chain Monte Carlo (MCMC) sampling process, collecting samples from random walks on the network of the known degree distribution, and using those samples to adjust the user's prior distribution into a posterior beta distribution. We consider SS-PSE's adjusted posterior beta distribution as our preferred estimate for the population size of Latino MSM who are HIV+ $Posterior N_{L,MSM,HIV+}$ and HIV unknown $Posterior N_{L,MSM,HIV?}$.

\subsection{Ramirez-Valles's Latino MSM Community Involvement}

Our primary data is the RDS network of individual respondents collected in the Latino MSM Community Involvement: HIV Protective Effects study \cite{ramirez2002protective}. This 2003-2004 survey features one of the original large-scale RDS sampling processes which collected a total of 643 samples of Latino MSM in the Chicago metro area (323 samples) and the San Francisco Bay area (320 samples). The study had several aims relevant to understanding the social determinants of HIV in the Latino MSM community, chief among them being whether a Latino MSM's sense of belonging or actions of involvement in their community had the so-called ``HIV Protective Effects." We take special interest in the use of this RDS data combined with prior knowledge from other sources to make population size estimates for Latino MSM with HIV in Cook and San Francisco County.



\subsection{American Community Survey's Same-Sex Households}

While Latino MSM and GBT-identifying populations are considered "hidden" from statistical researchers, Latino Male Same-Sex Householders (SSH) and their domestic partners can be readily identified within the American Community Survey (ACS). SSH populations across genders, ethnicities, years, and places can be studied by filtering the millions of individual ACS Public-Use Microdata household samples released every year on the variables of ``Sex" and ``Relationship with Householder." The U.S. Census Bureau has been producing reports on SSH populations this way since the ACS began in 2005 \cite{census3} \cite{census4}. 
\par
Interestingly, after a change in the graphic design of the survey form in 2008, there was a surprisingly significant drop in the number of SSH respondents across the nation. The U.S. Census Bureau concluded that the previous form resulted in respondents mischecking the box used to identify their sex. In 2011, after years of developing a statistical model relating a respondent's sex to their first name (first names and other personally identifying information are not included in the public data), the Census Bureau released their ``preferred" estimates for SSH demographics and population sizes for states across the U.S.

\begin{table*}[htbp]
  
  \centering
  \caption{Data sources and estimates leading a result for the number of Latino MSM who are HIV positive and HIV unknown using Respondent-Driven Sampling and county-level statistics.}
  \resizebox{\textwidth}{!}{
    \begin{tabular}{lllrlrl}
    \toprule
    Source & Symbol & Description & Cook County & 95\% & San Francisco County & 95\% \\
    \bottomrule \\
    
    \multirow{3}{*}{\cite{WilliamsInstitute}} & ${N_{SSH}}$ & \# of Same-Sex Householders, "SSH" & 14,050 & $\pm?$ & 10,450 & $\pm?$ \\ 
    
    & ${P_{M|SSH}}$ & \% SSH are Male & 68.33\% & $\pm?$ & 82.05\% & $\pm?$ \\ & ${P_{L|SSH}}$ & \% SSH are Latino & 12.55\% & $\pm?$ & 10.32\% & $\pm?$ \\ \\
    
    Eq. \eqref{ProbabilityOfLatinoMaleGivenSameSexHouseholder} & ${\hat{P}_{L,M|SSH}}$ & \% SSH are Latino and Male & 8.58\% & $\pm?$ & 8.47\% & $\pm?$ \\
    
    Eq. \eqref{NumberOfLatinoMaleSameSexHouseholder} & $\hat{N}_{L,M,SSH}$ & \# of Latino Male SSH & 1,205 & $\pm?$ & 885 & $\pm?$ \\
    \midrule \\
    
    \multirow{2}{*}{\cite{ramirez2013latino}} & $\hat{P}_{LivSit1|L,MSM}$ & \% Latino MSM living with partner only & 16.05\% & $\pm6.16\%$ & 14.79\% & $\pm5.71\%$ \\ 
    & $\hat{P}_{LivSit2|L,MSM}$ & \% Latino MSM living with partner and others & 3.14\% & $\pm2.98\%$ & 3.50\% & $\pm3.22\%$ \\ \\
    
    Eq. \eqref{ProbabilityThatLatinoMSMIsInDomesticPartnership} & $\hat{P}_{DomesPart|L,MSM}$ & \% Latino MSM are domestic partner & 19.19\% & $\pm6.84\%$ & 18.29\% & $\pm6.56\%$ \\
    
    Eq. \eqref{ProbabilityThatLatinoMSMIsHouseholder} & $\hat{P}_{SSH|L,MSM}$ & \% Latino MSM are SSH & 9.60\% & $\pm3.42\%$ & 9.15\% & $\pm3.28\%$ \\
    \midrule \\
    
    Eq. \eqref{PriorNumberOfLatinoMSM} & $Prior \hat{N}_{L,MSM}$ & \# of Latino MSM & 12,552 & $\pm6,946$ & 9,672 & $\pm5,405$ \\
        
    \midrule \\
    \multirow{2}{*}{\cite{census2}} & ${N_{L,M}}$ & Number of Latino Males & 462,801 & $\pm?$ & 54,251 & $\pm?$ \\
     & $P_{MSM|L,M}$ & \% Latino Male are MSM & 2.71\% & $\pm 1.50\%$ & 17.83\% & $\pm 9.96\%$ \\
    \midrule \\
    
    \multirow{2}{*}{\cite{ramirez2002protective}} & $\hat{P}_{HIV+|L,MSM}$ & \% Latino MSM are HIV positive & 14.1\% & $\pm6.85\%$ & 34.6\% & $\pm9.8\%$ \\ 
    & $\hat{P}_{HIV?|L,MSM}$ & \% Latino MSM are HIV unknown & 17.1\% & $\pm5.10\%$ & 10.1\% & $\pm6.86\%$ \\ \\
    
    \multirow{2}{*}{Eq. \eqref{NumberOfLatinoMSMHIV+}} & $\hat{N}_{L,MSM,HIV+}$ & \# of Latino MSM are HIV positive & 1,770 & $\pm860$ & 3,347 & $\pm948$ \\ 
    & $\hat{N}_{L,MSM,HIV?}$ & \# of Latino MSM are HIV unknown & 2,146 & $\pm640$ & 976 & $\pm657$ \\
    
    \bottomrule
    \end{tabular}%
    }
    \label{table_priors}
\end{table*}%

\begin{table*}[htbp]
  \centering
  \caption{Gile's SS weighted population estimates for a variety of living situations and HIV statuses in two counties.}
  \resizebox{\textwidth}{!}{
    \begin{tabular}{lrrrrrrrrrrrr}
    
    \toprule
    & \multicolumn{3}{l}{Cook County} & & & & \multicolumn{4}{l}{San Francisco County} & & \\
    \midrule
    
    Living Situation & \makecell[l]{Point\\Est.(\%)} & \makecell[r]{95\% \\Lower\\Bound} & \makecell[r]{95\% \\Upper\\Bound} & \makecell[r]{Est.\\Design\\Effect} & \makecell[r]{Std.\\ Error} & \makecell[r]{Sample\\Size} & \makecell[r]{Point\\Est.(\%)} & \makecell[r]{95\% \\Lower\\Bound} & \makecell[r]{95\% \\Upper\\ Bound} & \makecell[r]{Est.\\Design\\Effect} & \makecell[r]{Std.\\Error} & \makecell[r]{Sample\\Size} \\
    \midrule
    
    Alone in house / apartment & 25.27 & 17.86 & 32.65 & 3.44  & 3.77  & 84    & 22.30 & 13.48 & 31.13 & 3.01  & 4.50  & 59 \\
    Homeless & 0.89  & -0.25 & 2.03  & 1.76  & 0.58  & 3     & 3.67  & -0.24 & 7.60  & 2.91  & 2.00  & 6 \\
    School dormitory & 0.41  & -0.35 & 1.18  & 1.68  & 0.39  & 1     &       &       &      &      &       &  \\
    Residential hotel & 0.31  & -0.17 & 0.79  & 0.89  & 0.24  & 1     & 10.96 & 3.21  & 18.72 & 4.13  & 3.96  & 21 \\
    
    \makecell[l]{Shelter / halfway home / \\ \hspace{.5cm} rehabilitation facility} & 2.77  & -1.59 & 7.15  & 8.45  & 2.23  & 6     & 5.29  & 1.36  & 9.24  & 2.07  & 2.01  & 15 \\
    \makecell[l]{With domestic partner /\\ \hspace{.5cm} lover / boyfriend / others} & 3.14  & 0.16  & 6.13  & 3.49  & 1.52  & 10    & 3.50  & 0.28  & 6.70  & 2.04  & 1.64  & 9 \\
    With domestic partner / \\ \hspace{.5cm}lover / boyfriend & 16.05 & 9.89  & 22.18 & 3.34  & 3.14  & 45    & 14.79 & 9.08  & 20.45 & 1.72  & 2.90  & 41 \\
    \makecell[l]{With friend(s) or roommate(s)} & 25.55 & 18.96 & 32.20 & 2.74  & 3.38  & 92    & 33.30 & 24.35 & 42.25 & 2.42  & 4.57  & 84 \\
    
    With other relatives & 14.04 & 7.72  & 20.33 & 3.92  & 3.22  & 36    & 3.86  & 0.00  & 7.73  & 2.70  & 1.97  & 10 \\
    
    With parents or guardians & 11.57 & 6.00  & 17.18 & 3.64  & 2.85  & 36    & 2.33  & -0.42 & 5.07  & 2.22  & 1.40  & 5 \\
    \midrule     
    
    HIV-  & 68.80 & 60.55 & 77.17 & 3.83  & 4.24  & 205   & 55.30 & 46.41 & 64.31 & 2.17  & 4.57  & 142 \\
    
    HIV+  & 14.10 & 7.25  & 20.96 & 4.62  & 3.50  & 55    & 34.60 & 24.80 & 44.43 & 2.85  & 5.01  & 88 \\
    
    Unknown & 17.10 & 12.00 & 22.06 & 2.12  & 2.56  & 54    & 10.10 & 3.24  & 16.81 & 3.40  & 3.46  & 20 \\
    \bottomrule
    \end{tabular}%
    }
  \label{tab:PointEstimates}%
\end{table*}%

\begin{figure*}[t] \centering
\subfigure[An example of a $N_{L,MSM}$ Posterior for Cook County]{\includegraphics[width=8cm]{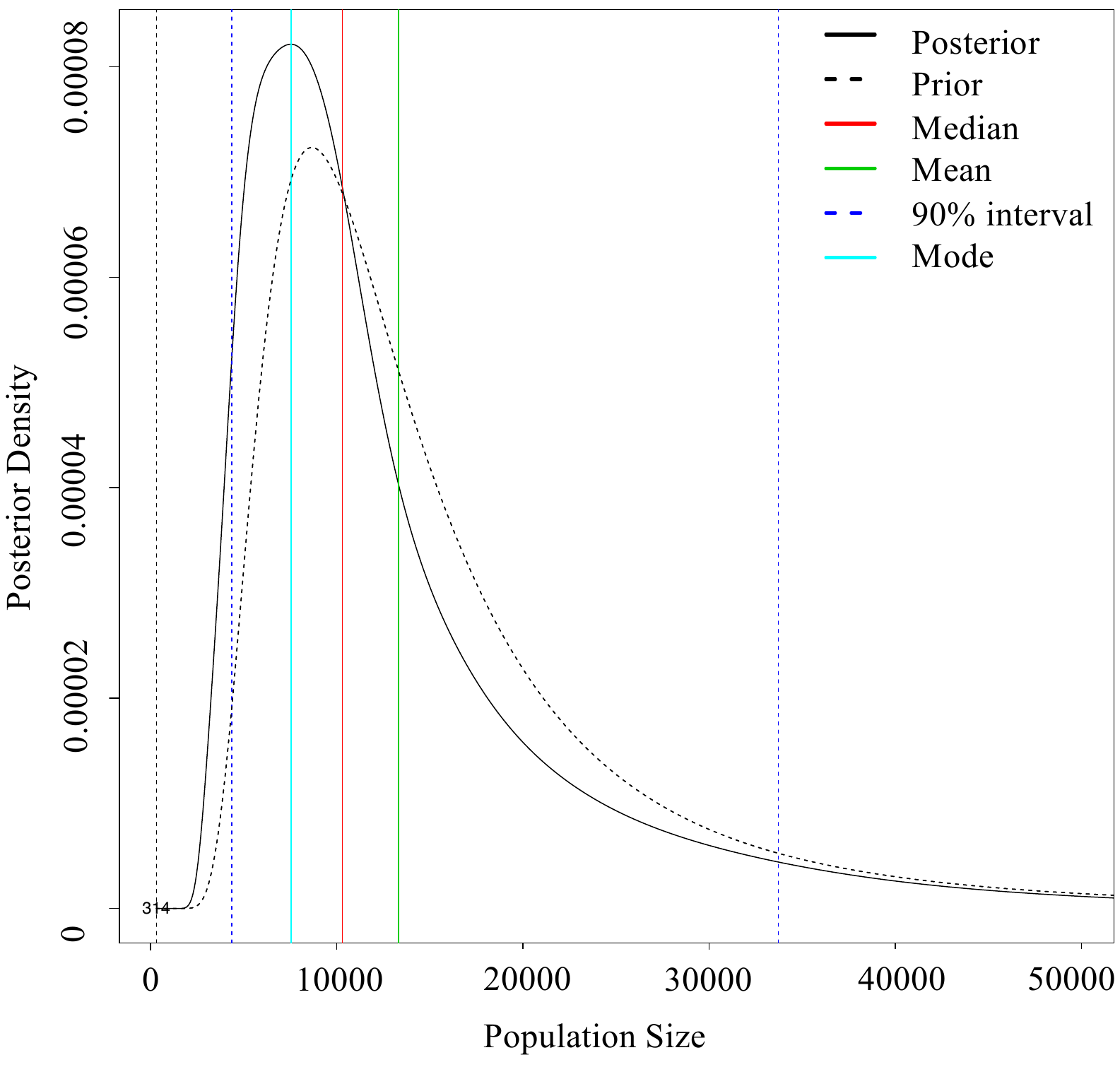}
\label{fig:CookCountyPosterior}}
\subfigure[An example of a $N_{L,MSM}$ Posterior for San Francisco County]
{\includegraphics[width=8cm]{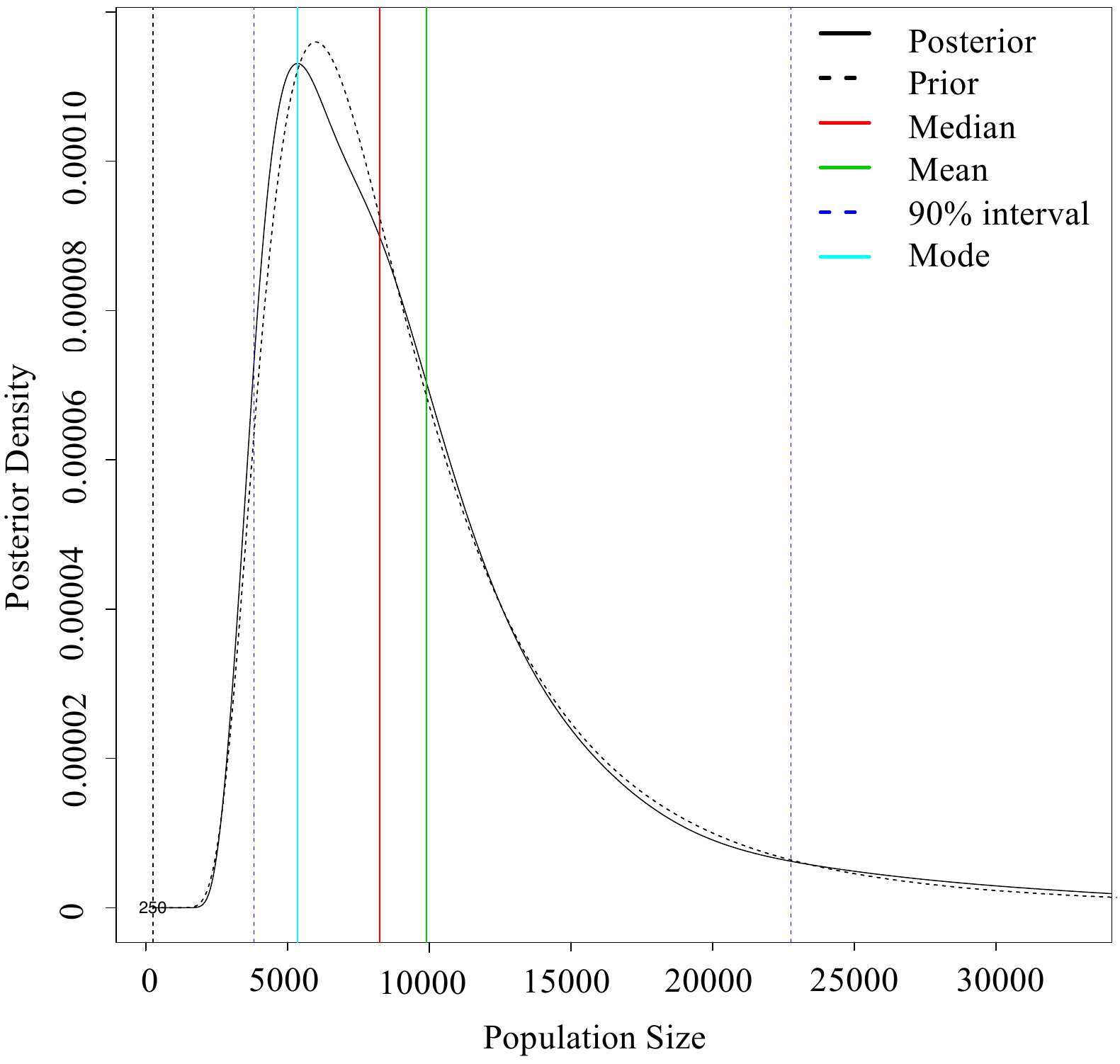}\label{fig:SanFranPosterior}
}
\caption{Example $N_{L,MSM}$ posteriors. The dashed "Prior" curves are the probability distributions fit by RDS Analyst to our prior $N_{L,MSM}$ estimates. The solid "Posterior" curves are the probability distributions on $N_{L,MSM}$ produced by SS-PSE.}
\label{fig:Posteriors}
\end{figure*}

\begin{table*}[htbp]

  \centering
  \caption{Summary of SS-PSE Posterior estimates. Priors from Table \ref{table_priors} are input into SS-PSE to estimate the mean $N_{L,MSM}$ across 10 trials. The resulting posterior estimates on $\hat{N}_{L,MSM,HIV+}$ and $\hat{N}_{L,MSM,HIV?}$ are re-applications of Eq. \eqref{NumberOfLatinoMSMHIV+} to those new estimates on $N_{L,MSM}$.}
    \begin{tabular}{lrrrrrrrr}
    \toprule
    \multicolumn{9}{c}{Cook County} \\
    \midrule &
    \makecell[r]{Mean} & 
    \makecell[r]{Median} &
    \makecell[r]{Mode} &
    \makecell[r]{25\%} & 
    \makecell[r]{75\%} & 
    \makecell[r]{90\%} & 
    \makecell[r]{2.5\%} & 
    \makecell[r]{97.5\%} \\
    \midrule
    
    $Prior \hat{N}_{L,MSM}$ & 12878 & 10708 & 7772  & 7660  & 15598 & 22654 & 4458  & 34801 \\
    $Posterior \hat{N}_{L,MSM}$ & 10071.5 & 8545.3 & 6904.6 & 6013  & 12335.4 & 17607.9 & 2583.9 & 27020.4 \\
    \midrule
    
    Relative Change & -21.79\% & -20.20\% & -11.16\% & -21.50\% & -20.92\% & -22.27\% & -42.04\% & -22.36\% \\
    
    
    \midrule
    $\hat{N}_{L,MSM,HIV+}$ & 1420.08 & 1204.89 & 973.55 & 847.83 & 1739.29 & 2482.71 & 364.33 & 3809.88 \\
    
    $\hat{N}_{L,MSM,HIV?}$ & 1722.23 & 1461.25 & 1180.69 & 1028.22 & 2109.35 & 3010.95 & 441.85 & 4620.49 \\
    \bottomrule
    
    \toprule
    \multicolumn{9}{c}{San Francisco County} \\
    \midrule
    
    $Prior \hat{N}_{L,MSM}$ & 9061  & 7722  & 5810  & 5642  & 10958 & 15475 & 3382  & 22986 \\
    $Posterior \hat{N}_{L,MSM}$ & 8469.8 & 7393.5 & 5925.8 & 5300.2 & 10416.5 & 14516.3 & 2448.5 & 21020.1 \\
    \midrule
    
    Relative Change & -6.52\% & -4.25\% & 1.99\% & -6.06\% & -4.94\% & -6.20\% & -27.60\% & -8.55\% \\
    
    \midrule
    
    $\hat{N}_{L,MSM,HIV+}$ & 2930.55 & 2558.15 & 2050.33 & 1833.87 & 3604.11 & 5022.64 & 847.18 & 7272.95 \\
    
    $\hat{N}_{L,MSM,HIV?}$ & 855.45 & 746.74 & 598.51 & 535.32 & 1052.07 & 1466.15 & 247.30 & 2123.03 \\
    
    \bottomrule
    \end{tabular}%
    \label{table_posteriors}
\end{table*}%

\subsection{William's Institute County-Level Estimates}

The Williams Institute's LGBT Data Interactive leverages the aforementioned state-level estimates in combination with their own model to present finer-grained county-level estimates for the number of Same-Sex Householders $N_{SSH}$, and the probability that the SSH is Latino $P_{L|SSH}$ or Male $P_{M|SSH}$ for all counties in the U.S. \cite{WilliamsInstitute}. 
\par 
We utilize their county-level estimates of Latino SSH and Male SSH and assume independence between $P_{L|SSH}$ and $P_{M|SSH}$ to produce an estimate of $P_{L,M|SSH}$. We combine that rate with their original estimate of overall population size $N_{SSH}$ to arrive at our prior estimate for the number of Latino Male Same-Sex Householders in each county, $N_{L,M,SSH}$. 


\begin{equation}
  \label{ProbabilityOfLatinoMaleGivenSameSexHouseholder}
    P_{M|SSH} \times P_{L|SSH} = \hat{P}_{L,M|SSH}
\end{equation}

\begin{equation}
  \label{NumberOfLatinoMaleSameSexHouseholder}
    \hat{P}_{L,M|SSH} \times N_{SSH} = \hat{N}_{L,M,SSH}
\end{equation}

\subsection{Latino MSM's Living Situation}

The question then is how to use the fairly reliable population size estimate of Latino Male Same-Sex Householders $\hat{N}_{L,M,SSH}$ to make an inference on the hidden Latino MSM population size, $\hat{N}_{L,MSM}$. 
\par
Our approach is to leverage Latino MSM answers to a single question in the Ramirez-Valles survey: ``Which of the following best describes your living situation?" Respondents who answered ``I am living with a domestic partner" $LivSit1$ or ``I am living with a domestic partner and other people" $LivSit2$ are aggregated into the single population of Latino MSM who are in a domestic partnership, $P_{DomesPart|L,MSM}$.

\begin{equation}
  \label{ProbabilityThatLatinoMSMIsInDomesticPartnership}
    \hat{P}_{LivSit1} + \hat{P}_{LivSit2} = \hat{P}_{DomesPart|L,MSM}
\end{equation}

\par
As the $LivSit$ question did not ask whether the respondent was the householder, we are left with the naive assumption that half of Latino MSM who are in a domestic partnership are also the householder, thus yielding a point estimate on $P_{SSH|L,MSM}$ - the critical ratio that bridges the known and hidden population size:

\begin{equation}
  \label{ProbabilityThatLatinoMSMIsHouseholder}
    \frac{1}{2} \times \hat{P}_{DomesPart|L,MSM} = \hat{P}_{SSH|L,MSM}
\end{equation}
\begin{equation}
  \label{PriorNumberOfLatinoMSM}
    \hat{N}_{L,M,SSH} / \hat{P}_{SSH|L,MSM} = \hat{N}_{L,MSM}
\end{equation}

In other words, the less likely Latino MSM are to be SSH, the greater our estimate for $\hat{N}_{L,MSM}$.

\subsection{Latino MSM's HIV Status}

The Ramirez-Valles dataset also provides observations on the ratio of Latino MSM who are HIV+ $\hat{P}_{HIV+|L,MSM}$ and HIV unknown $\hat{P}_{HIV?|L,MSM}$ in both Cook and San Francisco Counties. We obtained point estimates and 95\% confidence intervals for these variables using RDS Analyst.

\subsection{RDS Analyst's Population Estimates}
We use RDS Analyst to produce population proportion estimates using the Ramrirez-Valles RDS data weighted by Gile's Successive Sampling estimator (GSS) for the observed variables, $\hat{P}_{LivSit1}$ $\hat{P}_{LivSit2}$, $\hat{P}_{HIV+|L,MSM}$, and  $\hat{P}_{HIV?|L,MSM}$.

\subsection{RDS Analyst's SS-PSE}
We provided RDS Analyst's SS-PSE with a prior estimate on Latino MSM population size in the form of a 50\% confidence interval on $\hat{N}_{L,MSM}$. We obtained the 50\% confidence interval under the assumption that it was 34.4\% the width of our estimated 95\% confidence interval. We then run SS-PSE on that prior distribution for N = 10 trials and take the mean of the means to be our posterior point estimate on $\hat{N}_{L,MSM}$.


Finally, we use those posterior estimates on $\hat{N}_{L,MSM}$ and the same population proportion estimates to produce posterior estimates on the number Latino MSM who are HIV+ and unknown, $\hat{N}_{L,MSM,HIV+}$ and $\hat{N}_{L,MSM,HIV?}$.

\begin{equation}
  \label{NumberOfLatinoMSMHIV+}
    \hat{P}_{HIV|L,MSM} * \hat{N}_{L,MSM} = \hat{N}_{L,MSM,HIV}
\end{equation}



\section{Results}

As shown in the first section of Table \ref{table_priors}, the William's Institute's published results on the number of Same-Sex Householders, $\hat{N}_{L,M,SSH}$, in Cook and San Francisco County are 14,050 and 10,450, respectively \cite{WilliamsInstitute}. We could not find 95\% intervals for these estimates. 

The William's Institute also published estimates on the proportion of Latino SSH and Male SSH. Under the assumption of independence, we make an estimate on the proportion of Latino Male SSH for Cook and San Francisco counties to be 8.58\% and 8.47\%, respectively. The county proportion estimates are remarkably similar because, while San Francisco County has a larger Male SSH proportion, 82.05\% vs. 68\%, it has a lower Latino SSH proportion than Cook County, 10\% vs. 12.5\%. Multiplying the Latino Male SSH proportion by the SSH population, our final estimate for the number of Latino Male SSH is 1,205 and 885 in Cook and San Francisco county.

Using the 2004 LMSM-CI study\cite{ramirez2013latino} we make estimates driving towards the percentage of Latino MSM who are SSH. Table \ref{tab:PointEstimates} shows the categorical distribution across living situations for Latino MSM, with 95\% intervals computed by RDS Analyst. By taking the sum of two key living situations as shown in Eq. \eqref{ProbabilityThatLatinoMSMIsInDomesticPartnership}, we estimate the proportion of Latino MSM who are in a domestic partnership to be 19.19\% and 18.29\% with 95\% intervals $\pm \%6.85\%$ and $\pm 6.56\%$. Assuming that half of domestic partners are SSH leads to the critical estimate for proportion of Latino MSM who are SSH,$\hat{P}_{SSH|L,MSM}$ to be 9.60\% and 9.15\% with 95\% intervals $\pm 3.42\%$ and $\pm 3.28\%$, respectively. 

As shown in Eq. \eqref{PriorNumberOfLatinoMSM}, those ratios allow us to produce our prior estimates on the number of Latino MSM for Cook and San Francisco Counties to be 12,552 and 9,672 with 95\% intervals $\pm 6,946$ and $\pm 5,405$, respectively.


An interesting detail that can be derived from the calculations in Table \ref{table_priors} is the percentage of Latino males who are MSM. The number of Latino Males is provided by the U.S. Census Bureau's county intercensal estimates in 2011 \cite{census2}, as 462,801 and 54,251 for Cook and San Francisco County, respectively. The significantly smaller San Francisco Latino Male population implies a significantly larger probability that a Latino Male is a MSM in that county, with our estimates suggesting a probability of MSM given Latino Male to be about 1 in 6, 17.26\% $\pm 9.96\%$ in San Francisco County vs. 1 in 36, 2.75\% $\pm 1.5\%$ in Cook.


Table \ref{tab:PointEstimates} shows population estimates produced by RDS Analyst complete with 95\% intervals, sample sizes, and the estimated design effects. Using this table we can see that the sample sizes were close to zero for rarer living situations like ``Homeless" and ``School dormitory," a total of 9 Homeless and 1 School dormitory across both counties. Additionally, the rareness of a living situation varied between counties. In Cook County, Latino MSM are more likely to live with their parents, guardians or relatives, 25.61\% $\pm 5.57\%$ vs. San Francisco's relatively rare 6.19\% $\pm 4.73\%$. This may be an indication of a strong support network and familial roots for these men in Cook County vs. a population in San Francisco who tends to live in more non-traditional situations like residential hotels, shelters, or  homeless, with 19.92\% $\pm 3.95$ in San Francisco vs. 1.61\% $\pm 2.37$ in Cook.

To use SS-PSE we supply the prior estimates as calculated and displayed in Table \ref{table_priors}, specifically the Cook and San Francisco County Latino MSM population sizes, $N_{LMSM}$. We run SS-PSE 10 times for each county, summarizing the SS-PSE estimates obtained from those trials in Table \ref{table_posteriors}.

Each trial produces a beta distribution, an example of which is shown in Figure \ref{fig:Posteriors}. The dashed curve is the beta distribution fit by RDS Analyst to our prior estimates, which has been updated by SS-PSE to produce the posterior beta distribution shown as the solid curve. The vertical lines summarize the posterior beta distribution in terms of its mean (green), median (red), mode (light blue), and 90\% confidence interval (dark blue).

We use the mean of the mean estimates as the best point estimate for population size. The mean of the means of the 10 SS-PSE trials for Cook and San Francisco County, as shown in Table \ref{table_posteriors}, are 10,072 and 8,470 with standard errors 241 and 127, respectively. The 95\% bounds for the size of the Latino MSM population as computed by SS-PSE span from 2,584 to 27,020 for Cook County and 2,449 - 21,020 for San Francisco County.

Multiplying the same HIV proportions by the posterior estimates for the Latino MSM population size provided by SS-PSE, we arrive at our posterior estimates for the number of Latino MSM by HIV status in Cook and San Francisco counties to be 1,420 and 2,930 who are HIV positive and 1,722 and 855 who are HIV unknown, respectively.




\section{Conclusion}

Our method to estimate the total size of a hidden population featured several strengths and weaknesses. This is expected, as it is believed at this time that there is no single best method for estimating the size of a hard-to-reach population. This belief is based on work by Mauck et al. \cite{mauck2019population}, in which the authors review and compare multiple methods for estimating population size, but specifically for men who have sex with men. They determine that there is no single best method at this time and that in order to obtain a robust estimate, multiple methods should be used. 



In our study, we found that the methods available to us were expanded or limited by the inclusion of key variables in the RDS survey. We were fortunate to find a variable in our dataset which could be cross-referenced with data collected by the American Community Survey. This key variable, \textit{Living Situation}, provided evidence suggesting that the odds of a Latino MSM being a Same-Sex Householder is roughly 1 in 10 in both San Francisco and Cook County. The reliable estimates provided by the U.S. Census Bureau in the sampling of Same-Sex Householders combined with the consistency in the Living Situation distribution between the two counties is an encouraging sign for the existence of a statistically reliable relationship between SSH and MSM populations.

On the other hand, our methods were limited by the exclusion of key variables. For example, we could not use the generalized network scale-up method because our RDS dataset was missing one variable, ``Estimated Total Network Size." If our RDS survey included an estimation of the respondent's total network size along with their Latino MSM network size, we would have been capable of using NSUM as an additional technique to estimate the total Latino MSM population size.

We recommend future studies of MSM populations include a similar question to \textit{Living Situation}, and the collection of other variables which can be cross-referenced with the Same-Sex Householder populations studied by the American Community Survey. Doing so could enable the analysis of variance in those variables between Same-Sex Householders and MSM populations in counties across the U.S. Furthermore, as demonstrated in this paper, estimates on the proportion of MSMs who are SSH enables an additional method of population size estimation based on the reliable information on Same-Sex Householders as provided by the U.S. Census Bureau.

\balance

\bibliographystyle{ieeetr}
\bibliography{citations}

\end{document}